%% file: main.tex
\documentclass[letterpaper, 10 pt, conference]{ieeeconf}  

\IEEEoverridecommandlockouts                              

\usepackage{amsmath}

\usepackage{amsthm}
\usepackage{amssymb}
\usepackage{amsfonts}
\usepackage{graphicx}
\usepackage{cite}
\usepackage{dsfont}
\usepackage{mathrsfs}
\usepackage{siunitx}
\usepackage{subfigure}
\input{edits}
\newcommand{\real}[0]{\mathbb R}
\newcommand{\comp}[0]{\mathbb C}

\DeclareSymbolFont{bbold}{U}{bbold}{m}{n}
\DeclareSymbolFontAlphabet{\mathbbold}{bbold}

\newcommand{\diag}[1]{\ensuremath{\mathrm{diag}(#1)}}

\newtheoremstyle{bfnote}%
{}{}%
{\itshape}{}%
{\bfseries}{.}%
{ }%
{\thmname{#1}\thmnumber{ #2}\thmnote{ (#3)}}
\theoremstyle{bfnote}
\newtheorem{defi}{Definition}
\newtheorem{thm}{Theorem}

\newtheorem{rem}{Remark}

\newtheorem{ass}{Assumption}
\newtheorem{dyn-g}{Generator Dynamics}
\newtheorem{dyn-i}{Inverter Dynamics}

\title{\LARGE \bf
Grid-forming frequency shaping control
}

\author{Yan Jiang$^{1}$, Andrey Bernstein$^{2}$, Petr Vorobev$^{3}$, and Enrique Mallada$^{1}$
\thanks{This work was supported by any organization}
\thanks{$^{1}$Y. Jiang and E. Mallada are with the Johns Hopkins University, Baltimore, MD 21218, USA. Emails: {\tt \{yjiang,mallada\}@jhu.edu}}%
\thanks{$^{2}$A. Bernstein is with the National Renewable Energy Laboratory, Golden, CO 80401, USA.
        Email: {\tt andrey.bernstein@nrel.gov}}%
\thanks{$^{3}$P. Vorobev is with the Skolkovo Institute of Science and Technology, Moscow 143026, Russia.
        Email: {\tt P.Vorobev@skoltech.ru}}%
}

\begin{document}

\maketitle
\thispagestyle{empty}
\pagestyle{empty}

\begin{abstract} 
As power systems transit to a state of high renewable penetration, little or no presence of synchronous generators makes the prerequisite of well-regulated frequency for grid-following inverters unrealistic. Thus, there is a trend to resort to grid-forming inverters which set frequency directly. We propose a novel grid-forming frequency shaping control that is able to shape the aggregate system frequency dynamics into a first-order one with the desired steady-state frequency deviation and Rate of Change of Frequency (RoCoF) after a sudden power imbalance. The no overshoot property resulting from the first-order dynamics allows the system frequency to monotonically move towards its new steady-state without experiencing frequency Nadir, which largely improves frequency security. We prove that our grid-forming frequency-shaping control renders the system internally stable under mild assumptions. The performance of the proposed control is verified via numerical simulations on a modified Icelandic Power Network test case.

\end{abstract}


\section{Introduction}

Power system frequency control by storage units has been a topic of extensive research over the last decade, especially under the circumstances of the increasing penetration of renewable generation. Compared to conventional synchronous generators, storage units have outstanding ramping capabilities, which makes them an ideal choice for provision of various types of frequency control services. At present, special policies for storage participation in frequency control services are being developed by system operators around the world~\cite{hollinger2018fast,xu2016comparison}. For instance, the existing rules of the Enhanced Frequency Response --program introduced by National Grid in Great Britain-- already assume the power-frequency response with a gain of up to $100$ p.u.~\cite{greenwood2017frequency}, far exceeding typical capabilities of synchronous generators ($15$-$25$ p.u.). Thus, with the fall of the power system inertia and primary frequency reserves due to the increased penetration of renewables, energy storage systems have a potential to become the major providers of frequency control services in the future power systems.

So far, synthetic inertia and droop response by storage dominate the scientific literature. These two services are supposed to compensate for the falling system inertia and primary reserves, and seem to be a logical solution under existing grid codes. Typically, the storage units are supposed to realize the power-frequency type of response while being in the so-called grid-following mode. That is, inverters of the storage units measure the grid frequency and then inject (or consume) power based on a particular control strategy. Such an approach seems to be effective, yet 
the fact that there are certain delays associated with inverter control systems poses a threat to the frequency security. These delays are originated from the frequency measurement system  -- typically a phase-locked-loop (PLL), and also from inverter current control and pulse-width modulation (PWM) systems. It is foreseeable that, in the future low-inertia grid, these delays (from several decades of milliseconds to hundreds of milliseconds) can become fatal to frequency security. As an example, during the already famous South Australian blackout of 2016, the Rate of Change of Frequency (RoCoF) has hit the values as high as \SI{6}{\hertz\per\second}~\cite{yan2018anatomy}. Clearly, it becomes vital to develop new methods for storage participation in frequency control so as to minimize any possible response delays.

Grid-forming inverters \cite{pogaku2007modeling} have recently attracted a lot of attention from the research community, mainly in the context of autonomous microgrids. Beneficially, this type of inverters bring a broad range of new options for frequency control. First, they naturally adjust power almost with no delays (apart from some electro-magnetic transients in filters). Second, new control options become available. For instance, inertial response can be realized without any low-pass filters (hence, even less delays), since in the grid-forming mode this type of control becomes strictly causal. Third, inverters in the grid-forming mode are much less susceptible to grid voltage variations that often accompany frequency transients, which provides more reliability to the system. In the present manuscript, we explore a new approach for frequency control realized by grid-forming inverters -- a topic that is not yet studied sufficiently by both power and control communities. 

We propose a novel grid-forming \emph{frequency shaping control} that is inspired by its grid-following counterpart proposed in~\cite{jiang2020fs}. We first show that the proposed control is able to fashion the aggregate system frequency dynamics, a.k.a. Center of Inertia (CoI) Frequency, into a first-order one with the desired steady-state frequency deviation and RoCoF (following a sudden power imbalance). Notably, a first-order system frequency evolution naturally avoids overshoot so that the frequency deviation moves towards its steady-state incrementally without experiencing frequency Nadir, which is what we mean by ``Nadir elimination'' hereafter. Nadir elimination largely improves the frequency security since it reduces the risk of under-frequency load shedding. We then show that the proposed control ensures the internal stability of the overall system under mild conditions by using the decentralized stability criterion developed in~\cite{pm2019tcns}, where the crux of the matter is to check a positive realness (PR)~\cite{khalil2002nonlinear} requirement. We finally confirm the good performance of the proposed controller through numerical simulations on the modified Icelandic Power Network test case~\cite{iceland}.

\section{Power System Model}
\begin{figure}
\centering
\includegraphics[width=0.8\columnwidth]{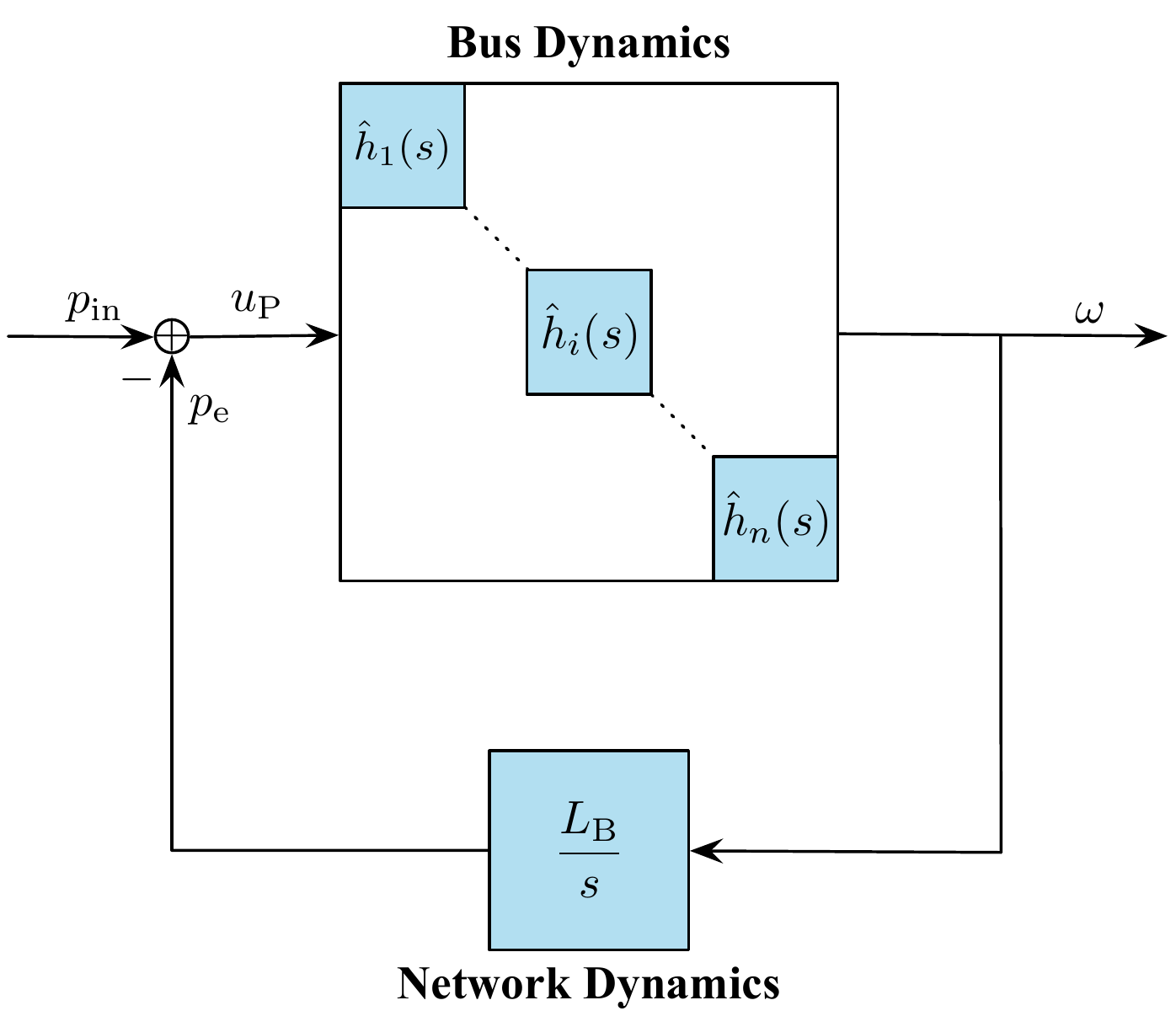}
\caption{Block diagram of power network.}\label{fig:model}
\end{figure}

We consider a power network composed of $n$ buses indexed by $i \in \mathcal{N} := \{1,\dots, n\} $ and transmission lines denoted by unordered pairs $\{i,j\} \in \mathcal{E}\subset \{\{i,j\}:i,j\in\mathcal{N},i\not=j\}$. As illustrated by the block diagram in Fig. \ref{fig:model}, the system dynamics are modeled as a feedback interconnection of bus dynamics and network dynamics.
The input signals $p_\mathrm{in} := \left(p_{\mathrm{in},i}, i \in \mathcal{N} \right) \in \real^n$ represent power injection changes and the output signals $\omega:=\left(\omega_i, i \in \mathcal{N} \right) \in \real^n$ represent the bus frequency deviations from its nominal value. We now discuss the dynamic elements in more detail.

\subsubsection{Bus Dynamics} 
The set of buses $\mathcal{N}$ is a disjoint union of the set of generator buses $\mathcal{G}$ and the set of inverter buses $\mathcal{I}$, i.e., $\mathcal{N}=\mathcal{G}\uplus\mathcal{I}$. The bus dynamics that map net power bus imbalances $u_\mathrm{P} := \left( u_{\mathrm{P},i}, i \in \mathcal{N} \right) \in \real^n$ to frequency deviations $\omega$ can be described by the transfer function matrix $\hat{H}(s) := \diag{\hat{h}_i(s), i \in \mathcal{N}}$, where $\hat{h}_i(s)$ is the transfer function of either generator or inverter depending on whether $i \in \mathcal{G}$ or $i \in \mathcal{I}$.
\paragraph{Generator Dynamics}
We consider generator dynamics that are composed of the standard swing dynamics with turbine droop, i.e.,
\begin{align} \label{eq:dy-sw-t}
\hat{h}_i(s) =& \left(m_i s + d_i + \frac{r_{\mathrm{t},i}^{-1}}{\tau_i s + 1}\right)^{-1}\,,\qquad \forall i \in \mathcal{G}\,,
\end{align}
where $m_i>0$ denotes the aggregate generator inertia, $d_i>0$ the aggregate generator damping, $\tau_i>0$ the turbine time constant, and $r_{\mathrm{t},i}>0$ the turbine droop coefficient.

\paragraph{Inverter Dynamics}
We consider grid-forming inverters, which set local grid frequency deviations $\omega_i$ directly as a function of their power output variation $q_{\mathrm{r},i} =-u_{\mathrm{P},i}$. The detailed function depends on the control law $\hat{h}_i(s)$ employed to map $u_{\mathrm{P},i}$ to $\omega_i$ for buses with $i \in \mathcal{I}$. 

\subsubsection{Network Dynamics}
The network power fluctuations $p_\mathrm{e} := \left(p_{\mathrm{e},i}, i \in \mathcal{N} \right) \in \real^n$ are given by a linearized model of the power flow equations~\cite{Purchala2005dc-flow}:
\begin{align}
 \hat p_\mathrm{e}(s) = \frac{L_\mathrm{B}}{s} \hat \omega(s)\;,\label{eq:N}
\end{align}
where $\hat p_\mathrm{e}(s)$ and $\hat \omega(s)$ denote the Laplace transforms of $p_\mathrm{e}$ and $\omega$, respectively.\footnote{We use hat to distinguish the Laplace transform from its time domain counterpart.} The matrix $L_\mathrm{B}$ is an undirected weighted Laplacian matrix of the network with elements 
\[
L_{\mathrm{B},{ij}}=\partial_{\theta_j}{\sum_{k=1}^n|V_i||V_k|b_{ik}\sin(\theta_i-\theta_k)}\Bigr|_{\theta=\theta_0}.
\]
Here, $\theta := \left(\theta_i, i \in \mathcal{N} \right) \in \real^n$ denotes the angle deviation from its nominal, $\theta_0 := \left(\theta_{0,i}, i \in \mathcal{N} \right) \in \real^n$ are the equilibrium angles, $|V_i|$ is the (constant) voltage magnitude at bus $i$, and $b_{ij}$ is the line $\{i,j\}$ susceptance.

\subsubsection{Closed-Loop Dynamics}
We are interested in the closed-loop response of the system in Fig.~\ref{fig:model} from the power injection changes $p_\mathrm{in}$ to frequency deviations $\omega$, which can be described by the transfer function matrix
\begin{equation}\label{eq:twp}
\hat{T}_{\omega \mathrm{p}}(s) :=\frac{\hat{\omega}(s)}{\hat{p}_\mathrm{in}(s)}= \left(I_n +\hat{H}(s)\frac{L_\mathrm{B}}{s}\right)^{-1} \hat{H}(s)\,. 
\end{equation}
 It is in general tough to analyze or tune the performance of $\hat{T}_{\omega \mathrm{p}}(s)$. Nevertheless, when the system is tightly-connected~\cite{Min2019cdc}, all buses exhibit a coherent response approximated by \begin{equation}\label{eq:sys-dyn}
   \hat{T}_{\omega \mathrm{p}}(s)\approx\hat{h}_\mathrm{c}(s) \mathbbold{1}_n \mathbbold{1}_n^T\,,
\end{equation}
where $\mathbbold{1}_n \in \real^n $ is the vector of all ones and 
\begin{equation}\label{eq:hc}
   \hat{h}_\mathrm{c}(s) :=\left(\sum_{i \in \mathcal{G}}\hat{h}^{-1}_i(s)+\sum_{i \in \mathcal{I}}\hat{h}^{-1}_i(s)\right)^{-1} \,.
\end{equation}
Henceforth, we refer to $\hat h_\mathrm{c}(s)$ in \eqref{eq:hc} as the \emph{coherent dynamics} of the network. 


\section{Grid-forming Frequency Shaping Control}\label{sec:design}
Motivated by \eqref{eq:sys-dyn}, we focus in this paper on shaping the response $\hat h_\mathrm{c}(s)$, instead of \eqref{eq:twp}. Thus, given generator dynamics $\hat{h}_i(s)$ for buses with $i \in \mathcal{G}$, our goal is to design inverter dynamics $\hat{h}_i(s)$ for buses with $i \in \mathcal{I}$ such that the coherent dynamics $\hat{h}_\mathrm{c}(s)$ is a first-order transfer function with two degrees of freedom. Such a coherent dynamics actually naturally ensures Nadir elimination as well as tunable steady-state frequency deviation and RoCoF, as the following theorem formally states.

\begin{thm}[Grid-forming frequency shaping control]\label{thm:GF-design}
Consider generator dynamics $\hat h_i(s)$, $i\in\mathcal{G}$, as in \eqref{eq:dy-sw-t}. Then, the grid-forming inverter control law
\begin{equation}\label{eq:hi-inverter}
\hat{h}_i(s) = \frac{1}{m_{\mathrm{I},i} s + d_{\mathrm{I},i}-\hat{g}_{\mathrm{I},i}(s)}\,,\qquad\forall i \in \mathcal{I}\,,
\end{equation}
with $m_{\mathrm{I},i}, d_{\mathrm{I},i}>0$, renders a first-order coherent dynamics
\begin{equation}\label{eq:hc-d}
    \hat{h}_\mathrm{c}(s) = \frac{1}{as+b}\,,
\end{equation}
with $a,b>0$ given by
\begin{subequations}\label{eq:I_parameters}
\begin{align}
 a :=\ &\sum_{i \in \mathcal{I}}m_{\mathrm{I},i} +\sum_{i \in \mathcal{G}} m_i\,,\label{eq:m-con}\\
b:=\ & \sum_{i \in \mathcal{I}} d_{\mathrm{I},i} +\sum_{i \in \mathcal{G}}d_i\,,
\end{align}
\end{subequations}
if and only if 
\begin{align}\label{eq:thm1-condition}
\sum_{i \in \mathcal{I}} \hat{g}_{\mathrm{I},i}(s) =& \sum_{i \in \mathcal{G}}\frac{r_{\mathrm{t},i}^{-1}}{\tau_i s + 1}\,.
\end{align}
In this case, the frequency deviations will experience no Nadir and the steady-state frequency deviations $\omega(\infty)$ and the RoCoF $|\dot{\omega}|_{\infty}$ will be determined by  
\begin{align}\label{eq:ab-tune}
    \omega(\infty)\!\approx\! \frac{\sum_{i=1}^n u_{0,i}}{b} \mathbbold{1}_n \;\;\;\text{and}\;\;\; {|\dot{\omega}|_{\infty} }\!\approx\! \frac{\sum_{i=1}^n u_{0,i}}{a} \mathbbold{1}_n \,,
\end{align}
when the system undergoes step power injection changes, i.e., $p_\mathrm{in} = u_0 \mathds{1}_{ t \geq 0 } \in \real^n$ with $u_0 \in \real^n$ being any arbitrary vector direction and $\mathds{1}_{ t \geq 0 }$ being the unit-step function.
\end{thm}
\begin{proof}
Applying the desired coherent dynamics given by \eqref{eq:hc-d} and the generator transfer function given by \eqref{eq:dy-sw-t} to the definition of coherent dynamics given by \eqref{eq:hc} yields 
\begin{equation*}
   as+b =\sum_{i \in \mathcal{G}}\left(m_i s + d_i+\frac{r_{\mathrm{t},i}^{-1}}{\tau_i s + 1}\right)+\sum_{i \in \mathcal{I}}\hat{h}^{-1}_i(s) \,.
\end{equation*}
Thus, the desired inverter control law should satisfy
\begin{equation*}
   \sum_{i \in \mathcal{I}}\hat{h}^{-1}_i(s) \!=\!\left(a-\!\sum_{i \in \mathcal{G}} m_i\right)s+\left(b-\!\sum_{i \in \mathcal{G}}d_i\right)-\sum_{i \in \mathcal{G}}\frac{r_{\mathrm{t},i}^{-1}}{\tau_i s + 1} \,.
\end{equation*}
It is straightforward that the control law determined by \eqref{eq:hi-inverter}, \eqref{eq:I_parameters}, and \eqref{eq:thm1-condition} guarantees that the above condition hold. This concludes the proof of the first statement. 

Next, combining \eqref{eq:twp} and \eqref{eq:sys-dyn}, we can see that the frequency deviations $\hat{\omega}(s)$ of the system $\hat{T}_{\omega \mathrm{p}}$ in response to step power injection changes $\hat{p}_\mathrm{in}(s) = u_0/s$ is given by
\begin{align}\label{eq:omega-hc}
    \hat{\omega}(s)=&\ \hat{T}_{\omega \mathrm{p}}(s)\hat{p}_\mathrm{in}(s)\approx\hat{h}_\mathrm{c}(s) \mathbbold{1}_n \mathbbold{1}_n^T\frac{u_0}{s}\nonumber\\=&\sum_{i=1}^n u_{0,i}\frac{\hat{h}_\mathrm{c}(s)}{s}\mathbbold{1}_n\,,
\end{align}
which can be interpreted as that the frequency deviation on each bus reacts to the aggregate step power injection change of size $\sum_{i=1}^n u_{0,i}$ with the coherent dynamics $\hat{h}_\mathrm{c}(s)$. Now, applying initial and final value theorems to \eqref{eq:omega-hc} with $\hat{h}_\mathrm{c}(s)$ given by \eqref{eq:hc-d}, we find that $a$ and $b$ satisfy the following relations:
\begin{subequations}
\begin{align*}
    {|\dot{\omega}|_{\infty} }\!=&\!\lim_{s\to\infty}\!s^2\hat{\omega}(s)\approx\!\!\lim_{s\to\infty} s^2\frac{\sum_{i=1}^n u_{0,i}}{s(as+b)}\mathbbold{1}_n
    \!=\!\frac{\sum_{i=1}^n u_{0,i}}{a}\mathbbold{1}_n\,,\\
    \omega(\infty)\!=&\lim_{s\to0}s\hat{\omega}(s)\approx \lim_{s\to0}s\frac{\sum_{i=1}^n u_{0,i}}{s(as+b)}\mathbbold{1}_n
    =\frac{\sum_{i=1}^n u_{0,i}}{b}\mathbbold{1}_n\,,
\end{align*}
\end{subequations}
which concludes the proof of \eqref{eq:ab-tune}. 
\end{proof}
Clearly, given specific requirements on steady-state frequency and RoCoF, there are infinite ways of choosing $m_{\mathrm{I},i}$ and $d_{\mathrm{I},i}$ to satisfy \eqref{eq:I_parameters}.
A straightforward choice is to set
\begin{align}
     m_{\mathrm{I},i}\! =\! \frac{a-\!\sum_{i \in \mathcal{G}} m_i}{|\mathcal{I}|}\ \text{and}\ d_{\mathrm{I},i}\! =\! \frac{b-\!\sum_{i \in \mathcal{G}}d_i}{|\mathcal{I}|}\,,\forall i \in\mathcal{I}\,,
\end{align}
where $|\mathcal{I}|$ denotes the cardinality of $\mathcal{I}$. 
Similarly, we propose the following two strategies to meet \eqref{eq:thm1-condition}.
\begin{itemize}
    \item \emph{Matching individual turbine dynamics by individual inverters:} Assume the cardinality of $\mathcal{I}$ is no less than that of $\mathcal{G}$, i.e., $|\mathcal{I}|\geq|\mathcal{G}|$. Let $\mathcal{I}_\mathrm{t}\subset\mathcal{I}$ such that there is a bijection between $\mathcal{I}_\mathrm{t}$ and $\mathcal{G}$ that maps each $j\in \mathcal{G}$ to distinct $i\in \mathcal{I}_\mathrm{t}$ by the following relation
    \begin{equation*}
    \hat{g}_{\mathrm{I},i}(s)=\frac{r_{\mathrm{t},j}^{-1}}{\tau_j s + 1}\,.
    \end{equation*}
    $\forall i \in \mathcal{I}\setminus \mathcal{I}_\mathrm{t}$, simply set $\hat{g}_{\mathrm{I},i}(s)=0$.
    \item \emph{Distributing the first-order reduced order model of the aggregate turbine dynamics~\cite{Min2019allerton} over inverters:} Let $z_i\geq0, \forall i \in\mathcal{I}$, be weighting parameters satisfying $\sum_{i \in\mathcal{I}} z_i =1$. Set
    \begin{equation*}
    \hat{g}_{\mathrm{I},i}(s)= z_i\frac{\tilde{r}_{\mathrm{t}}^{-1}}{\left(\tilde{\tau} s + 1\right)}\,,\qquad\forall i \in\mathcal{I}\,, \end{equation*}
    with $\tilde{r}_{\mathrm{t}}$ and $\tilde{\tau}$ being the turbine droop coefficient and time constant, respectively, of a first-order reduced order model of 
    \begin{equation*}
      \sum_{i \in \mathcal{G}}\frac{r_{\mathrm{t},i}^{-1}}{\tau_i s + 1}\,.  
    \end{equation*}
\end{itemize}
Tuning $\hat{g}_{\mathrm{I},i}(s)$ by distributing the first-order reduced order model of the aggregate turbine dynamics over inverters seems to be a more practical choice for two reasons. First, it gets rid of the need to accurately estimate droop coefficients and time constants of all individual turbines. Second, it relaxes the cardinality assumption $|\mathcal{I}|\geq|\mathcal{G}|$.

\begin{rem}[Meeting frequency specifications~\eqref{eq:ab-tune}] Choosing $a$ and $b$ to meet frequency specifications~\eqref{eq:ab-tune} naturally asks for knowledge of the current network composition via \eqref{eq:I_parameters} and \eqref{eq:thm1-condition}. The estimation of dynamic parameters, including but not limited to inertia, is currently an active research area~\cite{Phurailatpam2020tse, Schiffer2019tps, Min2019allerton}. This endorses our utilization of \eqref{eq:ab-tune} for safety specification. Arguably, whether \eqref{eq:I_parameters} holds rigorously for chosen $a$ and $b$ is not of major concern. We highlight that the proposed control always improve RoCoF for any positive $m_{\mathrm{I},i}$ and steady-state for large enough $d_{\mathrm{I},i}$, $\forall i \in\mathcal{I}$. 
\end{rem}

\begin{rem}[Steady-state power output from grid-forming frequency shaping control inverters] It is easy to show from \eqref{eq:dy-sw-t}, \eqref{eq:hc}, \eqref{eq:hi-inverter}, and \eqref{eq:I_parameters} that the steady-state power output from the proposed inverters depends on the relation between $d_i$ for $i \in \mathcal{I}$ and $r_{\mathrm{t},i}^{-1}$ for $i \in \mathcal{G}$. Note that, if $\mathcal{I}=\emptyset$, then $\hat{h}_\mathrm{c}(0)=1/\sum_{i \in \mathcal{G}} \left(d_i + r_{\mathrm{t},i}^{-1}\right)$; otherwise $\hat{h}_\mathrm{c}(0)=1/\left( \sum_{i \in \mathcal{G}} d_i + \sum_{i \in \mathcal{I}} d_{\mathrm{I},i}\right)$. Hence, as long as $\sum_{i \in \mathcal{I}} d_{\mathrm{I},i}>\sum_{i \in \mathcal{G}}r_{\mathrm{t},i}^{-1}$, the collection of inverters will provide power in steady-state since the steady-state frequency deviation will be reduced. 

\end{rem}
\begin{rem}[Freedom of resources allocation]
The coherent dynamics $\hat h_c(s)$ depends merely on the summation of the inverse of grid-forming frequency shaping control transfer functions $\hat{h}_i(s)$ over $i\in\mathcal{I}$, but not on the way of how these  control resources are distributed across the network. Although, in our discussion above,  control resources are mainly equally distributed over inverters, there are actually many other possibilities. 
Thus, a promising future research direction will be the exploration of how to optimally allocate control resources based on additional performance metrics that may be of interest.
\end{rem}

Considering the two choices of $\hat{g}_{\mathrm{I},i}(s)$ suggested before, we make the following assumption on the form of $\hat{g}_{\mathrm{I},i}(s)$.

\begin{ass}[The form of $\hat{g}_{\mathrm{I},i}(s)$]\label{ass:gi}
$\forall i \in \mathcal{I}$, $\hat{g}_{\mathrm{I},i}(s)$ is in one of the two forms below, i.e., 
\begin{equation}\label{eq:gi-form}
  \hat{g}_{\mathrm{I},i}(s)=0\qquad\text{or}\qquad\hat{g}_{\mathrm{I},i}(s)=\frac{\rho_i}{\sigma_i s + 1} \,,  \end{equation}
  where $\rho_i,\sigma>0$.
\end{ass}




\section{Stability Analysis}
In this section, we show that the grid-forming frequency shaping control given by \eqref{eq:hi-inverter} and \eqref{eq:gi-form} ensures internal stability of the overall system in Fig.~\ref{fig:model} under mild conditions compatible with \eqref{eq:thm1-condition}. 
To this end, we first review some standard concepts that play a role in our stability analysis.
\begin{defi}[$\mathcal{H}_\infty$ space~\cite{Zhou1996robust}] $\mathcal{H}_\infty$ is the Hardy space of functions $\hat{F}(s)$ that are analytic in the open right-half complex plane $\comp_+$ with a bounded norm $\|\hat{F}\|_\infty := \sup_{s\in \comp_+} |\hat{F}(s)|$.
\end{defi}

\begin{defi}[Positive real~\cite{khalil2002nonlinear}] A proper rational transfer function matrix $\hat{F}(s)$ is called positive real (PR) if:
\begin{itemize}
\item Poles of all elements of $\hat{F}(s)$ are in the closed left-half complex plane $\overline{\comp}_-$.
\item For any $\nu \in \real$ such that $\boldsymbol{j}\nu$ is not a pole of any element of $\hat{F}(s)$, the matrix $\hat{F}(\boldsymbol{j}\nu)+\hat{F}^T(-\boldsymbol{j}\nu)$ is positive semidefinite.
\item For any $\nu \in \real$ such that $\boldsymbol{j}\nu$ is a pole of some element of $\hat{F}(s)$, the pole $\boldsymbol{j}\nu$ is simple and the residue matrix $\lim_{s\to\boldsymbol{j}\nu}\left(s-\boldsymbol{j}\nu\right)\hat{F}(s)$ is positive semidefinite Hermitian.
\end{itemize}
Here, $\boldsymbol{j}$ represents the imaginary unit that satisfies $\boldsymbol{j}^2=-1$.
\end{defi}

\begin{rem}[Real rational subspace of $\mathcal{H}_\infty$]
The real rational subspace of $\mathcal{H}_\infty$ consists of all proper real rational stable transfer matrices. Thus, in order to check whether a proper real rational transfer function belongs to $\mathcal{H}_\infty$ or not, it is sufficient to check whether it is stable or not.
\end{rem}

\begin{rem}[Applications of positive realness]
The positive realness was originally introduced in electrical network synthesis~\cite{otto1931thesis} and recently extended to mechanical network synthesis~\cite{Smith2002tac}. Moreover, it has been applied a lot to stability analysis for both linear and nonlinear systems.
\end{rem}

We are now ready to conduct a stability analysis.

\begin{thm}[Internal stability under grid-forming frequency shaping control]\label{thm:stable}
Let Assumption~\ref{ass:gi} hold. The system $\hat{T}_{\omega \mathrm{p}}$ with \eqref{eq:dy-sw-t} and \eqref{eq:hi-inverter} is internally stable if $d_{\mathrm{I},i} > \rho_i$, $\forall i\in \mathcal{I}$ with nonzero $\hat{g}_{\mathrm{I},i}(s)$. 
\end{thm}

\begin{proof}
According to the decentralized stability criterion proposed in \cite{pm2019tcns}, the system $\hat{T}_{\omega \mathrm{p}}$ is internally stable if $\exists \tau_\alpha, \epsilon>0$ such that
\begin{align}\label{eq:stable-con}
     \gamma_i\hat{h}_i(s)\in\mathcal{Q}\,,\qquad\forall i\in \mathcal{N}\,,
\end{align}
with
\begin{align}
    \mathcal{Q}\!&:=\!\left\{\!\hat{q}(s)\in\mathcal{H}_\infty\left|\  \hat{q}(0)\neq 0, \frac{s}{s+\tau_\alpha}\!\!\left(1\!+\!\frac{\hat{q}(s)}{s}\right)\!\!-\!\epsilon\! \in \!\text{PR}\!\right.\right\}\!\!\,,\nonumber\\
    \gamma_i\!&:=2\sum_{j=1}^n \overline{V}_i \overline{V}_j b_{ij}\,,\nonumber
\end{align}
where $\overline{V}_i$ and $\overline{V}_j$ denote the maximum allowable voltage magnitudes at endpoints of the line $\{i,j\}$. Thus, the key is to check whether the condition in \eqref{eq:stable-con} holds for $\hat{h}_i(s)$, $\forall i\in \mathcal{N}$. 


Combining \eqref{eq:hi-inverter} and \eqref{eq:gi-form}, we know that $\forall i \in \mathcal{I}$,
\begin{align*}
    \hat{h}_i(s)\!=\!\frac{1}{m_{\mathrm{I},i} s + d_{\mathrm{I},i}}\ \text{or}\ \hat{h}_i(s) \!=\!\left( m_{\mathrm{I},i} s + d_{\mathrm{I},i}\!-\!\frac{\rho_i}{\sigma_i s + 1}\right)^{-1}\!\!\!\!\,.
\end{align*}

We begin with the later case, from which we get
\begin{align}\label{eq:hi-2nd}
    \gamma_i \hat{h}_i(s)=\frac{ \gamma_i\left(\sigma_i s + 1\right) }{m_{\mathrm{I},i} \sigma_i s^2 + \left(m_{\mathrm{I},i} + d_{\mathrm{I},i} \sigma_i \right) s + d_{\mathrm{I},i} - \rho_i}\,.
\end{align}
First, it is well-known that a second-order transfer function is stable if all coefficients of its denominator have the same sign. Thus, $m_{\mathrm{I},i}, d_{\mathrm{I},i}, \sigma_i>0$, and $d_{\mathrm{I},i} > \rho_i$, $\forall i\in \mathcal{I}$, guarantee the stability of \eqref{eq:hi-2nd}, i.e., $\gamma_i \hat{h}_i(s) \in \mathcal{H}_\infty$. Second, it is trivial to check that $\gamma_i \hat{h}_i(0) =\gamma_i/\left(d_{\mathrm{I},i} - \rho_i\right)\neq0$. Last but not least, we need to show that $\exists \tau_\alpha, \epsilon>0$ such that
\begin{align*}
    \frac{1}{s+\tau_\alpha}\!\!\left[s+\frac{ \gamma_i\left(\sigma_i s + 1\right) }{m_{\mathrm{I},i} \sigma_i s^2 \!+\! \left(m_{\mathrm{I},i} + d_{\mathrm{I},i} \sigma_i \right) s \!+\! d_{\mathrm{I},i} \!-\! \rho_i}\right]\!\!-\!\epsilon\!\in \!\text{PR}\,,
\end{align*}
which is equivalent to 
\begin{align}\label{eq:hi-PR-con}
    \frac{\xi_{3,i} s^3 + \xi_{2,i} s^2 + \xi_{1,i} s + \xi_{0,i}}{\eta_{3,i} s^3 + \eta_{2,i} s^2 + \eta_{1,i} s + \eta_{0,i}} \in \text{PR}
\end{align}
with 
\begin{subequations}\label{eq:xi}
\begin{align}
    &\xi_{0,i}\!:=\gamma_i -\left(d_{\mathrm{I},i} - \rho_i\right)\tau_\alpha\epsilon \,,\\ &\xi_{1,i}\!:=\left(d_{\mathrm{I},i} \!-\! \rho_i\right)\left(1\!-\!\epsilon\right)+\gamma_i\sigma_i-\left(m_{\mathrm{I},i}\! +\! d_{\mathrm{I},i} \sigma_i \right) \tau_\alpha\epsilon\,,\\ &\xi_{2,i}:=\left(m_{\mathrm{I},i} + d_{\mathrm{I},i} \sigma_i \right)\left(1-\epsilon\right)- m_{\mathrm{I},i} \sigma_i \tau_\alpha \epsilon\,,\\
    &\xi_{3,i}:=m_{\mathrm{I},i} \sigma_i\left(1-\epsilon\right)\,,\\
    &\eta_{0,i}:=\left(d_{\mathrm{I},i} - \rho_i\right)\tau_\alpha\,,\\
    &\eta_{1,i}:=\left(d_{\mathrm{I},i} - \rho_i\right)+\left(m_{\mathrm{I},i} + d_{\mathrm{I},i} \sigma_i \right) \tau_\alpha\,,\\
    &\eta_{2,i}:=m_{\mathrm{I},i} + d_{\mathrm{I},i} \sigma_i + m_{\mathrm{I},i} \sigma_i \tau_\alpha\,,\\
    &\eta_{3,i}:= m_{\mathrm{I},i} \sigma_i\,.
\end{align}
\end{subequations}
We now show that \eqref{eq:hi-PR-con} holds by performing the algebraic test for positive realness proposed in~\cite{Chen2009tac}. That is, for the nondegenerate case, i.e., $\left(\xi_{0,i},\xi_{1,i},\xi_{2,i},\xi_{3,i}  \right)^T \in \real_{\geq 0}^{4}$ and $\left(\eta_{0,i},\eta_{1,i},\eta_{2,i},\eta_{3,i}  \right)^T \in \real_{\geq 0}^{4}\setminus \mathbbold{0}_4$ with $\mathbbold{0}_4$ being the zero vector of size $4$, the condition \eqref{eq:hi-PR-con} holds if and only if
\begin{equation}\label{eq:3rd-pr-test}
 \left(\xi_{1,i}+\eta_{1,i}\right) \left(\xi_{2,i}+\eta_{2,i}\right)  \geq   
 \left(\xi_{0,i}+\eta_{0,i}\right) \left(\xi_{3,i}+\eta_{3,i}\right)\,.
\end{equation}
We check the nonnegativity of all coefficients in \eqref{eq:hi-PR-con} first. Suppose $\tau_\alpha>0$ and $0<\epsilon<1$. Clearly, it follows directly from $m_{\mathrm{I},i}, d_{\mathrm{I},i}, \sigma_i>0$, and $d_{\mathrm{I},i} > \rho_i$, $\forall i\in \mathcal{I}$, that $\xi_{3,i}, \eta_{0,i},\eta_{1,i},\eta_{2,i},\eta_{3,i}>0$. Also, for any given $\tau_\alpha>0$, $\xi_{0,i},\xi_{1,i},\xi_{2,i}>0$ if $\epsilon$ is sufficiently small. Now we are ready to check whether \eqref{eq:3rd-pr-test} holds or not. Applying \eqref{eq:xi} to the left hand side of \eqref{eq:3rd-pr-test} yields
\begin{align}\label{eq:3rd-pr-test-lhs}
 &\left(\xi_{1,i}+\eta_{1,i}\right) \left(\xi_{2,i}+\eta_{2,i}\right)  \\=&\left[\left(d_{\mathrm{I},i} - \rho_i\right)\left(2-\epsilon\right)+\gamma_i\sigma_i+\left(m_{\mathrm{I},i} + d_{\mathrm{I},i} \sigma_i \right)\tau_\alpha\left(1-\epsilon\right)\right]\nonumber\\
 &\left[\left(m_{\mathrm{I},i} + d_{\mathrm{I},i} \sigma_i \right)\left(2-\epsilon\right)+ m_{\mathrm{I},i} \sigma_i \tau_\alpha \left(1-\epsilon\right) \right]\,.\nonumber
\end{align}
Applying \eqref{eq:xi} to the right hand side of \eqref{eq:3rd-pr-test} yields
\begin{align}\label{eq:3rd-pr-test-rhs}
  &\left(\xi_{0,i}+\eta_{0,i}\right) \left(\xi_{3,i}+\eta_{3,i}\right)\\
  =&\left[\gamma_i +\left(d_{\mathrm{I},i} - \rho_i\right)\tau_\alpha\left(1-\epsilon\right)\right] m_{\mathrm{I},i} \sigma_i\left(2-\epsilon\right)\,.\nonumber
\end{align}
Through standard algebra, using \eqref{eq:3rd-pr-test-lhs} and \eqref{eq:3rd-pr-test-rhs}, we get \begin{align*}
 &\left(\xi_{1,i}+\eta_{1,i}\right) \left(\xi_{2,i}+\eta_{2,i}\right)- \left(\xi_{0,i}+\eta_{0,i}\right) \left(\xi_{3,i}+\eta_{3,i}\right) \\
 =&\left(d_{\mathrm{I},i} - \rho_i\right)\left(m_{\mathrm{I},i} + d_{\mathrm{I},i} \sigma_i \right)\left(2-\epsilon\right)^2\\&+\left(m_{\mathrm{I},i} + d_{\mathrm{I},i} \sigma_i \right)^2\tau_\alpha\left(2-\epsilon\right)\left(1-\epsilon\right)
 +\gamma_i d_{\mathrm{I},i} \sigma_i^2 \left(2-\epsilon\right)\nonumber\\
 &+\left[\gamma_i\sigma_i+\left(m_{\mathrm{I},i} + d_{\mathrm{I},i} \sigma_i \right)\tau_\alpha\left(1-\epsilon\right)\right]m_{\mathrm{I},i} \sigma_i \tau_\alpha \left(1-\epsilon\right)\nonumber\\
 \geq&\ 0\,,
\end{align*}
for any sufficiently small $\epsilon$, which means \eqref{eq:3rd-pr-test} holds. Thus, the required positive realness in \eqref{eq:hi-PR-con} has been proved. Therefore, $\gamma_i\hat{h}_i(s)\in\mathcal{Q}$ in this case.

We then turn to the simple case where
\begin{align}\label{eq:hi-1st}
    \gamma_i\hat{h}_i(s)=\frac{\gamma_i}{m_{\mathrm{I},i} s + d_{\mathrm{I},i}}\,.
    \end{align}
First, the stability of \eqref{eq:hi-1st}, i.e., $\gamma_i \hat{h}_i(s) \in \mathcal{H}_\infty$, follows from the fact that the only pole of it is $-d_{\mathrm{I},i}/m_{\mathrm{I},i}<0$. Second, $\gamma_i \hat{h}_i(0) =\gamma_i/d_{\mathrm{I},i} \neq0$. As for the required positive realness, \eqref{eq:hi-1st} can be considered as a special case of \eqref{eq:hi-2nd} with $\rho_i=0$ and $\sigma_i=0$. Plugging $\rho_i=0$ and $\sigma_i=0$ into \eqref{eq:xi} gives $\xi_{0,i},\xi_{1,i},\xi_{2,i}, \eta_{0,i},\eta_{1,i},\eta_{2,i}>0$, $\xi_{3,i}=\eta_{3,i}=0$, and 
\begin{align*}
 &\left(\xi_{1,i}+\eta_{1,i}\right) \left(\xi_{2,i}+\eta_{2,i}\right)- \left(\xi_{0,i}+\eta_{0,i}\right) \left(\xi_{3,i}+\eta_{3,i}\right) \\
=&\ d_{\mathrm{I},i}m_{\mathrm{I},i} \left(2-\epsilon\right)^2+m_{\mathrm{I},i} ^2\tau_\alpha\left(2-\epsilon\right)\left(1-\epsilon\right)\geq 0\,,
\end{align*}
for any sufficiently small $\epsilon$, which lead to the required positive realness. Therefore, $\gamma_i\hat{h}_i(s)\in\mathcal{Q}$ in this case. 

Finally, from \eqref{eq:dy-sw-t}, we know that $\forall i \in \mathcal{G}$,
\begin{align}\label{eq:hi-ge}
    \gamma_i \hat{h}_i(s)=\frac{ \gamma_i\left(\tau_i s + 1\right) }{m_i \tau_i s^2 + \left(m_i + d_i \tau_i \right) s + d_i + r_{\mathrm{t},i}^{-1}}\,.
\end{align}
Observe that \eqref{eq:hi-ge} and \eqref{eq:hi-2nd} have the same form except for some minor sign differences. Thus, the proof of $\gamma_i\hat{h}_i(s)\in\mathcal{Q}$ follows from a similar argument on \eqref{eq:hi-2nd}.
This concludes the proof that the system $\hat{T}_{\omega \mathrm{p}}$ is internally stable. 
\end{proof}


\section{Numerical Illustrations}
In this section, we present simulation results that compare the novel grid-forming frequency shaping control with the popular grid-forming virtual inertia control~\cite{Poolla2019place}. The simulations are conducted on the Icelandic Power Network available in the Power Systems Test Case Archive \cite{iceland}. Instead of the linearized network model used in the analysis, the simulations are built upon a nonlinear setup including nonlinear power flows and line losses. The original dynamic model contains $35$ generator buses and $83$ load buses, whose union is denoted as $\mathcal{N}$. To mimic a low-inertia scenario, we only keep $6$ generator buses that are equipped with turbines out of original $35$ generator buses. Each of above $6$ generator buses is distinctly indexed by some $i\in\left\{1,\ldots,6\right\}:=\mathcal{G}$ here. We then randomly pick $6$ buses from the set $\mathcal{N}\setminus\mathcal{G}$ as inverter buses. Each of above $6$ inverter buses is distinctly indexed by some $i\in\left\{7,\ldots,12\right\}:=\mathcal{I}$ here. The remaining buses are left as load buses denoted by $\mathcal{L}:=\mathcal{N}\setminus\left(\mathcal{G}\cup\mathcal{I}\right)$.

For every generator bus $i\in\mathcal{G}$, the aggregate generator inertia $m_i$, the turbine time constant $\tau_i$, and the turbine droop coefficient $r_{\mathrm{t},i}$ are directly obtained from the dataset. In addition, turbine governor deadbands are taken into account such that turbines are only responsive to frequency deviations exceeding \SI{\pm0.036}{\hertz} \cite{vorobev2019deadbands}. Given that the values of generator damping coefficients are not provided by the dataset, we set $d_i = 1\ \text{p.u.}$. For every load buses $i\in\mathcal{L}$, the damping coefficient is chosen as $1/20$ of the mean of all generator damping coefficients, i.e., $\bar{d} := (\sum_{i\in\mathcal{G}} d_i)/|\mathcal{G}|$.

The inverter control law on buses $i\in\mathcal{I}$ is either grid-forming virtual inertia (GF-VI) or grid-forming frequency shaping (GF-FS). The GF-VI is modelled as
\begin{equation*}\label{eq:hi-GF-VI}
\hat{h}_i(s) = \frac{1}{m_{\mathrm{v},i} s + d_{\mathrm{v},i}}\,,\qquad\forall i \in \mathcal{I}\,,
\end{equation*}
where $m_{\mathrm{v},i}>0$ is the virtual inertia constant and $d_{\mathrm{v},i}>0$ is the virtual damping constant. $\forall i \in \mathcal{I}$, we set $m_{\mathrm{v},i}=\bar{m}:= (\sum_{i\in\mathcal{G}} m_i)/|\mathcal{G}|$ and $d_{\mathrm{v},i}=\bar{d}$. As for the GF-FS in \eqref{eq:hi-inverter}, we only test the more practical tuning method suggested in Section~\ref{sec:design}, where $\hat{g}_{\mathrm{I},i}(s)$ is obtained by distributing the first-order reduced model of the aggregate turbine dynamics over inverters. Thus, $\forall i \in \mathcal{I}$, we set $m_{\mathrm{I},i}=\bar{m}$,  
\begin{equation*}
    d_{\mathrm{I},i}=\bar{d}+\frac{\tilde{r}_{\mathrm{t}}^{-1}}{6}\qquad\text{and}\qquad\hat{g}_{\mathrm{I},i}(s)=\frac{\tilde{r}_{\mathrm{t}}^{-1}}{6\left(\tilde{\tau} s + 1\right)}\,,
\end{equation*}
which ensures that the RoCoF and steady-state frequency deviations under GF-VI and GF-FS are the same so as to provide a fair comparison. Note that, with this setting, the stability condition required in Theorem~\ref{thm:stable} is satisfied since $d_{\mathrm{I},i}= \bar{d}+\tilde{r}_{\mathrm{t}}^{-1}/6> \tilde{r}_{\mathrm{t}}^{-1}/6=\rho_i$, $\forall i\in \mathcal{I}$.

\begin{figure}[t!]
\centering
\subfigure[System without inverters, where inverters on buses $i\in \mathcal{I}$ are replaced by loads with damping coefficients given by $\bar{d}/20$ and the generator damping is increased so as to   exactly compensate the lost inverter damping]
{\includegraphics[width=\columnwidth]{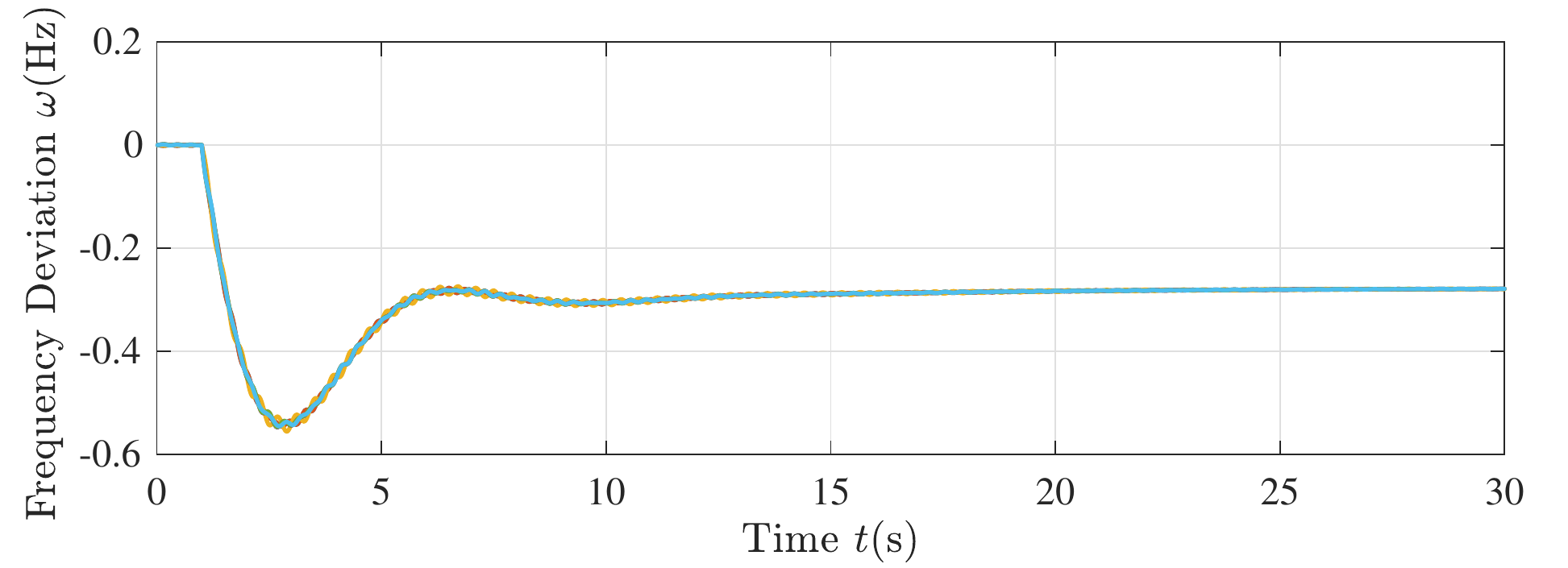}\label{fig:sw-simu}}
\hfil
\subfigure[System with GF-VI inverters]
{\includegraphics[width=\columnwidth]{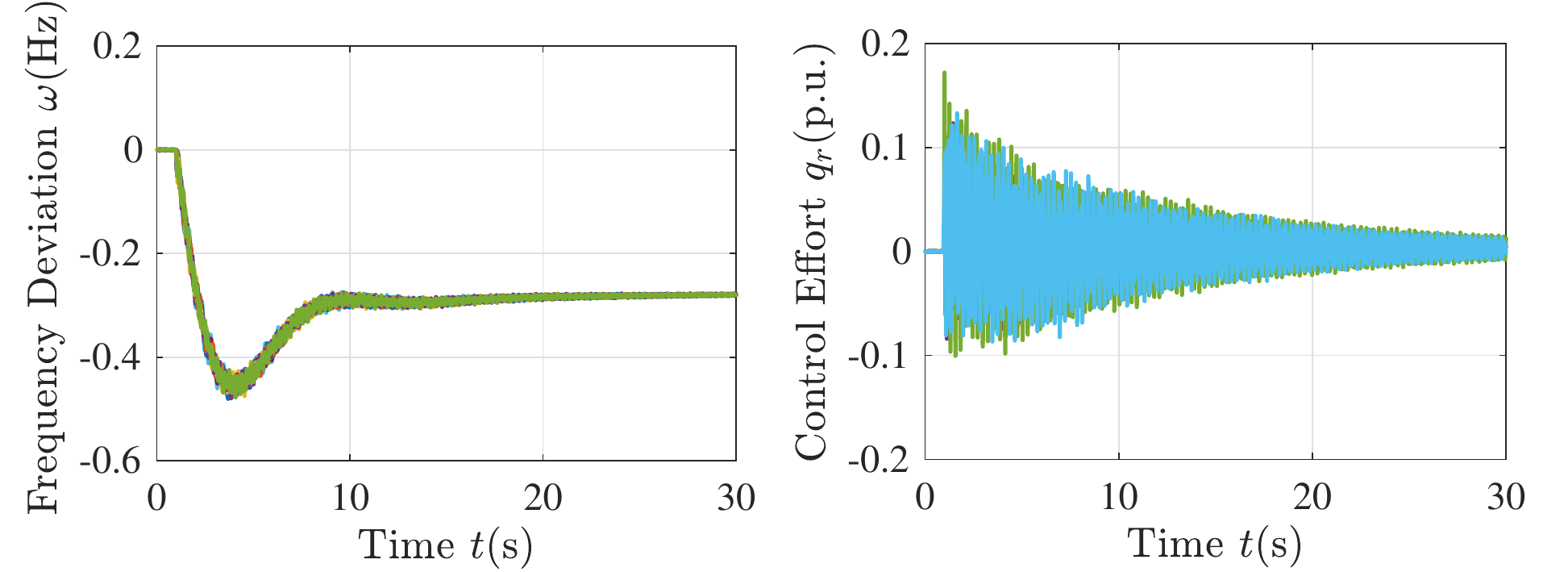}\label{fig:GF_VI}}
\hfil
\subfigure[System with GF-FS inverters]
{\includegraphics[width=\columnwidth]{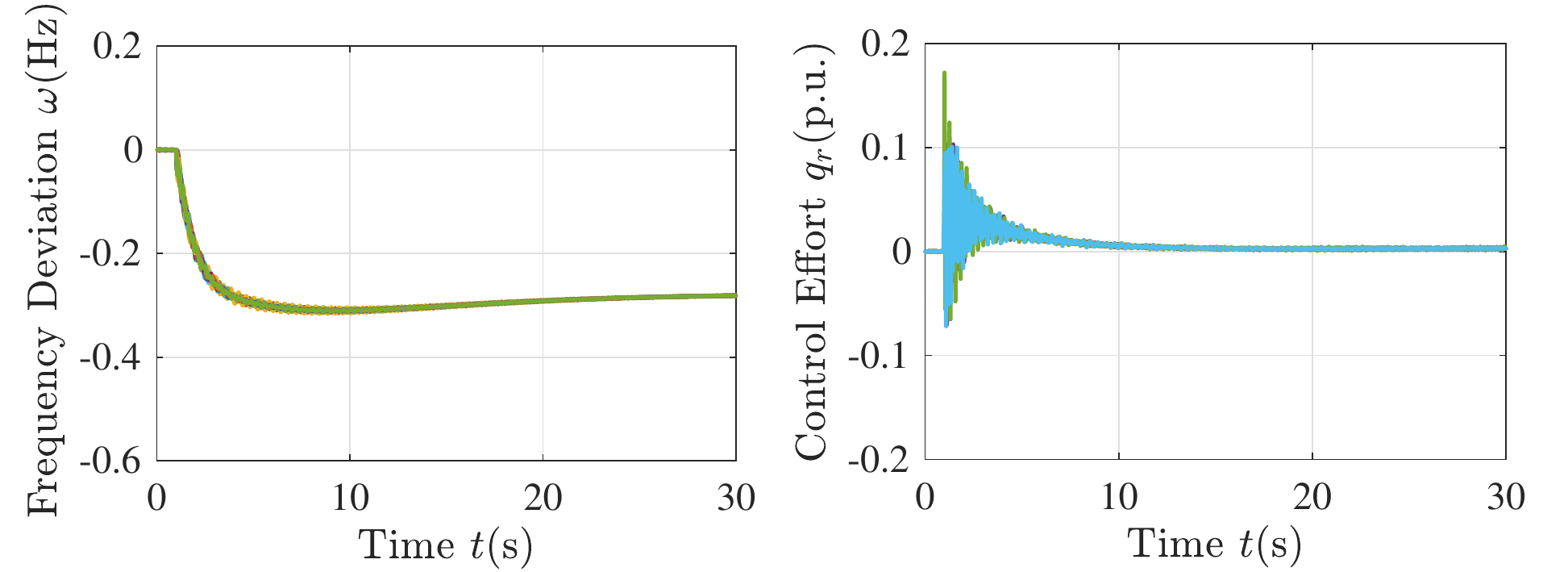}\label{fig:GF_FS_method2}}
\caption{Performance of the system when a $-0.3$ p.u. step change in power injection is introduced to a randomly picked bus.}
\label{fig:compareS}
\end{figure}

For the purpose of comparison, the frequency deviation of the system without inverters when there is a step change of $-0.3$ p.u. in power injection at a randomly picked bus at time $t = \SI{1}{\second}$ is provided in Fig.~\ref{fig:sw-simu}. The performances of the system under the two inverter control laws  are given in
Fig.~\ref{fig:GF_VI} and Fig.~\ref{fig:GF_FS_method2}. Some observations can be made. First, the system under GF-FS almost exhibits a first-order coherent dynamics as predicted by Theorem~\ref{thm:GF-design}, while the system under GF-VI experiences a deep Nadir. Second, Nadir elimination via GF-FS only requires an acceptable amount of control effort.

\section{Conclusions and Future Work}
A novel grid-forming frequency shaping control has been proposed for inverter-based frequency control in low-inertia power systems. The proposed control is able to force the system frequency to exhibit first-order coherent dynamics with specified steady-state frequency deviations and RoCoF in response to sudden power injection changes. The key benefit of a first-order frequency response is that the frequency deviations gradually evolve towards the final equilibrium  without experiencing Nadir so as to improve frequency security. The internal stability of the system is guaranteed by the proposed control under mild conditions. The performance of the proposed control is verified through numerical simulations. 

Future work include: (i) developing a more advanced control to achieve a second-order coherent dynamics with desired steady-state frequency deviations, RoCoF, and tunable Nadir; (ii) investigating the problem of optimal allocation of the proposed control resources over the network; (iii) considering a more detailed inverter model to throw light to device-level execution of the proposed control.

\addtolength{\textheight}{-12cm}   







\bibliographystyle{IEEEtran}
\bibliography{GFFS}

\end{document}

%% file: edits.tex
\usepackage{ifthen}
\newboolean{showcomments}
\setboolean{showcomments}{true}
\usepackage[usenames,dvipsnames]{color}
\usepackage{todonotes}

\definecolor{bleudefrance}{rgb}{0.19, 0.55, 0.91}
\definecolor{ao(english)}{rgb}{0.0, 0.5, 0.0}

\newcommand{\addcite}[0]{\ifthenelse{\boolean{showcomments}}
{\textcolor{purple}{(add cite(s)) }}{}}%

\newcommand{\enrique}[1]{  \ifthenelse{\boolean{showcomments}}
{\todo[inline,color=bleudefrance]{Enrique: #1}}{}}
\newcommand{\emmargin}[1]{\ifthenelse{\boolean{showcomments}}{\marginpar{\color{bleudefrance}\tiny EM: #1}}{}}

\newboolean{showedits}
\setboolean{showedits}{false}
\usepackage[markup=underlined]{changes}
\definechangesauthor[color=bleudefrance]{EM}
\newcommand{\aem}[1]{
\ifthenelse{\boolean{showedits}}
{\added[id=EM]{#1}}
{\!#1\hspace{-4.75pt}}
}
\newcommand{\repem}[2]{
\ifthenelse{\boolean{showedits}}
{\replaced[id=EM]{#1}{#2}}
{\!#1\hspace{-4.75pt}}
}
\newcommand{\dem}[1]{
\ifthenelse{\boolean{showedits}}
{\deleted[id=EM]{#1}}
{}
}